\documentclass[nofootinbib,showkeys]{revtex4}
\usepackage{color}
\usepackage{array}
\usepackage{float}
\usepackage{multirow}
\usepackage{amsmath}
\usepackage{amssymb}
\usepackage{graphicx}
\usepackage[english]{babel}

\usepackage{amsfonts}
\usepackage{amssymb}
\usepackage{amsthm}
\usepackage{indentfirst}
\usepackage[utf8]{luainputenc}
\usepackage{varioref}
\labelformat{equation}{\textrm{(#1)}}
\labelformat{thm}{Teorema {#1}}

\theoremstyle{plain}

\theoremstyle{definition}

\newtheorem{defn*}{Definizione}

\theoremstyle{remark}

\allowdisplaybreaks
\newtheorem{proo}{\textbf{Dimostrazione}}{\nonumber}

\begin{document}

\title{Capillary Network, Cancer and Kleiber Law}


%\author{G. Dattoli\thanksref{e2,addr2}
%        \and
%        E. Di Palma\thanksref{addr2}.
%}
\author{G. Dattoli}
\email{giuseppe.dattoli@enea.it}

\author{E. Di Palma }
\email{emanuele.dipalma@enea.it}

\author{S. Licciardi }
\email{silviakant@gmail.com}
\affiliation{ENEA - Centro Ricerche Frascati, Via Enrico Fermi 45, 00044, Frascati, Rome, Italy}

\author{C. Guiot}
\email{caterina.guiot@unito.it}
\affiliation{ University of Turin, Department of Surgical Science, Via C. Raffaello 30, 10124 Torino, Italy }

\author{T.S. Deisboeck }
\email{deisboec@helix.mgh.harvard.edu -- ts.deisboeck@thinkmotu.com}
\affiliation{ Department of Radiology, Harvard Medical School, Massachusetts General Hospital 149 13th St Charlestown, MA, 02129}

%\thankstext{addr1}{e-mail: giuseppe.dattoli@enea.it}
%\thankstext{addr1}{e-mail: emanuele.dipalma@enea.it}
%\thankstext{addr2}{e-mail: emanuele.dipalma@enea.it}

%\affiliation{%INFN- Laboratori Nazionali di Frascati,Via Enrico Fermi, 00044, Frascati, Rome, Italy \label{addr1}
%           %\and
%           ENEA - Centro Richerche Frascati, Via Enrico Fermi 45, 00044, Frascati, Rome, Italy \label{addr2} }
%           %\and
%           %\emph{Present Address:} if needed\label{addr3}
%}


\begin{abstract}
We develop a heuristic  model embedding Kleiber and Murray laws to describe mass growth, metastasis and vascularization in cancer. We analyze the relevant dynamics using different evolution equations (Verhulst, Gompertz and others).  Their extension to reaction diffusion equation of the Fisher type is then used to describe the relevant metastatic spreading in space. Regarding this last point, we suggest that cancer diffusion
%[we need to be careful with the terms: ‘expansion’ is used to describe growth of the primary tumor, ‘invasion’ is spatial migration of (non-proliferating) cancer cells, and ‘dissemination’ is used for metastasis – please chose]
 may be regulated by Levy flights mechanisms and discuss the possibility that the associated reaction diffusion equations are of the fractional type, with the fractional coefficient being determined by the fractal nature of the capillary evolution. \end{abstract}


\maketitle

\section{Introduction}
%\addcontentsline{toc}{chapter}{Introduzione}{}
%\markboth{\textsc{Introduction}}{}

The Kleiber law \cite{Kleiber} has been successfully used by West, Brown and Enquist (WBE) \cite{WBE} to derive, starting from energetic balance considerations, a differential equation yielding the evolution of biological masses.
 Such a point of view has been applied to the growth of cancer masses \cite{Guiot}, which, like any biological evolving complex, follows the logistic paradigm contained in the WBE equation, namely exponential growth and saturation.
  The use of WBE, with suitable modifications, has also provided further insights into aspects of cancer metabolism \cite{Dattoli1} and metastatis \cite{Dattoli2}.\\

The approach developed in \cite{Guiot,Dattoli2} does not make any explicit reference to cancer vascularization, which is the autonomous capillary network that feeds its entire mass \cite{Carmeliet}.
 It is generally accepted that the cancer origin be monoclonal and that the tumor passes through a transition phase called dormancy, in which the mutated cells take up nutrients from the surrounding, without having developed any autonomous capillary architecture yet.\\

Apparently, there is not a specific time regulating the duration of this phase, which may last indefinitely \cite{Ghiso}. However when nutrients flow to cancer tissues through an ad hoc developed capillary web, the associated mass starts growing exponentially \cite{Folkman}.\\

The critical radius of the cancer mass (supposed spherical in shape), before reaching the angio-genetic stage, can be assumed to be around $0.1-0.2$ $mm$. The associated threshold (critical) mass, assuming the tumor density to be close to that of water, is on the order of $m_{T}\backsimeq 30\cdot 10^{-6}g$.
 Since the mass of the single constituent cells can be taken to be $m_{c}\backsimeq 3\cdot10^{-9}g$  the number of cells in the latent cancer complex is approximatly $10^{4}$.\\

Although there is a general consensus with regards to the existence of agents promoting cancer vascularization, there is less agreement on what sequence precisely triggers the angio-genic response \cite{Ghiso}. In this paper we do not discuss these mechanisms, rather we consider the problem in dynamical and energetic terms, by noting that in the pre-angio-genetic (avascular) phase the cancer mass fluctuates around the mass threshold value. We will therefore model this first part using the na\"{\i}ve equation
\begin{equation}\begin{split}\label{mPA}
& \dfrac{d}{dt}m_{PA}=\lambda-\gamma m_{PA}\\
& m_{PA}(0)=0, \qquad m_{T}=\dfrac{\lambda}{\gamma}
\end{split}\end{equation}
the threshold mass $m_{T}$ is obtained after setting $\dfrac{d}{dt}m_{PA}=0$.\\

 If we assume that the life span of a cancer cell be some days (we assume $10$ to get round numbers), we can infer that $\lambda\left[\dfrac{g}{day}\right]\simeq 3\cdot 10^{-6} $ , which amounts to a production rate of $10^{3}$  cells per day. By taking into account that the energy necessary to generate a single cell is $2.1\cdot 10^{-5}$ $J$ \cite{WBE}, we can calculate that the power necessary to sustain the cell rate production of a dormant tumor is around $0.2$ $\mu W$ .\\

This limited amount of power is just enough to sustain the metabolic rate of the cancer complex, that forms the threshold mass. It cannot drive the system to the sizeable levels of grams. An idea of the power demand necessary to accomplish these tasks is offered by the use of an allometric relation of the Kleiber type. We adapt
indeed the $\frac{3}{4}$
% \footnote{In the following we will discuss the deep meaning of allometric “laws” and in particular the Kleiber law, which for the moment will be assumed as an empirical relationship between power demand of a biological system and its mass.}{\color{red}{??}}
  power law of the living systems to cancer by expressing the power necessary to sustain a tumor of mass $m$ as
 \begin{equation}\begin{split}
 & P(m)= Bm^{\frac{3}{4}}\\
 & B\simeq 3.4\cdot 10^{-2}\frac{W}{g^{\frac{3}{4}}}
\end{split} \end{equation}
According to eq. \ref{mPA} we obtain, in correspondence of $m_{T}$, a power demand in the order of $P_{T}\simeq 14$ $\mu W$. We can now estimate the amount of oxygen needed to sustain this cellular threshold mass. At the maximum of saturation $1$ litre of blood contains $0.2$ litres of oxygen \cite{Westerhof}, therefore we have
\begin{equation}
l_{O_{2}}\simeq 0.2 l_{b}
\end{equation}
The amount of power linked to the oxygen consumption rate is
\begin{equation}
P_{O_{2}}[W]\simeq 2\cdot 10^{4} \dfrac{\Delta l_{O_{2}}}{\Delta t}\left[ \dfrac{l}{s}\right]
\end{equation}
We find therefore from the previous equations that a tumor of mass $m$  requires the following blood flow
\begin{equation}
\dfrac{\Delta l_{b}}{\Delta t}\left[ \dfrac{l}{s}\right]\simeq 2.5 \cdot10^{-4}Bm^{\frac{3}{4}}
\end{equation}
which, for the previously calculated threshold mass corresponds to a blood flux rate of about $3.5 \dfrac{nl}{s}\simeq 3.5 \dfrac{\mu g}{s}$.\\

The aforementioned threshold value of $14$ $\mu W$ represents the maximum power demand which the dormant mass can draw from the surrounding environment without the need of an autonomous vessel system. Above this value, hypoxia may set in and will act as the main promoter of the secretion of angiogenetic signals.\\

In the forthcoming sections we will see how the above concepts can be embedded into a dynamical model of cancer mass growth, including the development of the capillary network system.

\section{Mass and Capillary Web Growth}

Before going further we consider the problem from a merely “hydraulic” point of view and remind the reader that the blood is viewed as a Newtonian fluid of constant viscosity and the flow is assumed to be Poiseuille \cite{Westerhof}, the vessels are considered rigid tubes and the pressure gradient is assumed to be constant.  According to Poisseille law \cite{Westerhof}, transmural pressure and blood flux are related by the following relation, which mimics the Ohm law
\begin{equation}\begin{split}\label{resist}
& p=RQ\\
& R=\dfrac{8\eta L}{\pi r^{4}}
\end{split}\end{equation}

In the previous equations $p$  is the transmural pressure, $R$  is the hydraulic resistance and $Q$ is the volumetric flow. The analogy with the Ohm law is evident and it is more than a mere similitude.\\

The resistance is linked to the length of the vessel, $L$, to its radius $r$ and to the blood viscosity $\eta$.\\

 The purely “hydraulic” power, dissipated to maintain a constant blood flux in the capillary network is
 \begin{equation}
 P_{H}=RQ^{2}
 \end{equation}
The total dissipated power needs a further contribution from the metabolic term, which provides the energy necessary to the tissues to survival \cite{Westerhof}, namely
\begin{equation}\begin{split}
& P_{M}=\alpha_{b}\pi r^{2}L\\
&\alpha_{b}\left[\dfrac{W}{l}\right]\simeq 2\cdot 10^{4}
\end{split}\end{equation}
 with $\alpha _{b}$ being the metabolic rate per unit volume of blood.\\

 Murray's law states that the vascular system is organized in such a way that a balance exists between this last term and the power required for blood circulation \cite{Westerhof}. The volumetric blood flow should be maximized to provide nutrients to tissues, but an increase of $Q$ implies a correspondingly (quadratic)  increase of the dissipated hydraulic power to sustain the blood flow rate. Murray’s law is based on a minimization principle for the dissipated power, $P$. The assumptions of Newtonian fluid behavior and the constancy of the flow gradient allow the following formulation of the ''design'' principle for a single vessel of length $L$ and radius $r$:
\begin{equation}\begin{split}
& \dfrac{\partial P}{\partial r}=0\\
& P=P_{H}+P_{M}
\end{split}\end{equation}
The total power is minimized whenever
\begin{equation}\label{minPot}
Q=\dfrac{\pi r^{3}}{4}\sqrt{\dfrac{\alpha _{b}}{\eta}}
\end{equation}
The essence of Murray's law is contained in eq. \ref{minPot}; furthermore, owing to the mass continuity at a capillary bifurcation we have $Q_{0}=Q_{1}+Q_{2}$ and therefore $r_{0}^{3}=r_{1}^{3}+r_{2}^{3}$  . Assuming for now that at each bifurcation the daughter capillaries have the same radius, we find that, after n-separation, the n-th capillary radius is $r_{n}=2^{-\frac{n}{3}}r_{0}$.\\

We can safely assume that initially, namely at the beginning of the vascularization process, the capillary network is just consist of a single element feeding the tumor mass, therefore
\begin{equation}
2\alpha_{b}\pi r^{2}L=BM^{\frac{3}{4}}
\end{equation}
Assuming that $L$  be of the order of the cancer mass radius, assumed to be spherical, we find
\begin{equation}\begin{split}
& r^{2}=\dfrac{BM^{\frac{3}{4}}}{2\pi \alpha_{b}L}\\
& L=\left(\dfrac{3M}{4\pi \rho} \right)^{\frac{1}{3}}
\end{split}\end{equation}
accordingly, this yields an initial vessel radius of the order of $(20-30)\mu m$.\\

We expect therefore that capillary network formation is promoted from the outside and starts penetrating inside the tumor mass by diffusion and branching. A very insightful description of the process can be found in ref. \cite{Wise}.\\

% Fig. \ref{angiogen} yields an idea of how capillary network evolves inside the tumor mass.
%\begin{figure}[h]
% \centering
% \includegraphics[width=.5\textwidth]{angiogen}
% \caption{\textbf{Tumor and Capillary Network} }\label{angiogen}
% \end{figure}

According to the previous discussion, while the capillaries diffuse and branch inside the tumor we expect that their radius is reduced as predicted by Murray's law, furthermore their length is expected to scale as the radius of the capillary itself. Assuming that $\dfrac{L_{n}}{r_{n}}=\dfrac{L_{0}}{r_{0}}$ , we find that the total metabolic power delivered to the tumor mass is
\begin{equation}\label{potenza}
P_{M}^{T}=\alpha_{b}\pi \sum _{n=0}^{N}2^{n}r_{n}^{2}L_{n}=N\alpha_{b}\pi r_{0}^{2}L_{0}
\end{equation}
since $2^{n}r_{n}^{2}L_{n}=r_{0}^{2}L_{0}$. Eq. \ref{potenza} captures the essence of Murray's law, n the sense that the parameters combine so that all capillaries contribute to the tumor mass feeding as a single capillary of length $NL_{0}$.\\

We can now try to understand what could be the temporal evolution of $N$, which should be related to the tumor radius.\\

After the dormant phase, when the capillary network emerges, the solution of the WBE equation yields the following mass evolution
\begin{equation}\label{WBE}
m(t)=M_{\infty}\left[1-\left(1-\sqrt[4]{\dfrac{m_{0}}{M_{\infty}}} \right)e^{-\frac{t}{\tau_{1}}}  \right]^{4}
\end{equation}
Where  $M_{\infty},\tau_{1}$ are the “saturated” mass and the characteristic time of the evolution, respectively. We have denoted by $m_{0}$ the initial tumor mass, which is taken to be equivalent to $m_{T}$. Either $M$ or $\tau_{1}$ can be expressed in terms of the constants of the theory, but for the moment they will be assumed to be free parameters eventually derived from the experimental data.\\

Assuming the tumor spherical in shape, we can derive from eq. \ref{WBE} the time dependence of the radius, namely
\begin{equation}\label{raggio}
r(t)=R_{\infty}\left[1-\left(1-\sqrt[4]{\dfrac{m_{0}}{M_{\infty}}} \right)e^{-\frac{t}{\tau_{1}}} \right]^{\frac{4}{3}}
\end{equation}
Evolution of tumor mass $(gram)$ and radius $(m)$ are reported in Fig. \ref{m(t)r(t)}; for the case of
$M_{\infty}\simeq 800g$, $\tau _{1}\simeq 107days$, it is worth noting that within the present picture, whenever $m_{0}<<M_{\infty}$, $\tau_{1}$ is the time necessary to reach $16\%$ of the final mass.\\\\

%\begin{table}
%\begin{figure} %[h]
%\begin{minipage}[b]{0.47\textwidth}
% \centering
%%\begin{tabular}{cc}
% \includegraphics[width=.6\textwidth]{m(t).pdf}
% \caption{\label{m(t)} $a)$ Tumor mass $(g)$ vs. time $(days)$ \qquad \qquad
%  $M_{\infty}\simeq 800g$, $\tau_{1}\simeq 107$ $days$}
%\end{minipage}
%\hfill
%\begin{minipage}[b]{0.47\textwidth}
% %\end{figure}
%%\begin{figure}[h]
% \centering
% \includegraphics[width=.6\textwidth]{r(t).pdf}
% \caption{\label{r(t)} $b)$	Corresponding Tumor radius $(m)$ vs. time $(days) $ $M_{\infty}\simeq 800g$, $\tau_{1}\simeq 107$ $days$}
%\end{minipage}
% \end{figure}
%\end{tabular}
%\end{table}
\begin{figure} %[h]
 \begin{minipage}[b]{0.47\textwidth}
 \centering
 \includegraphics[width=.6\textwidth]{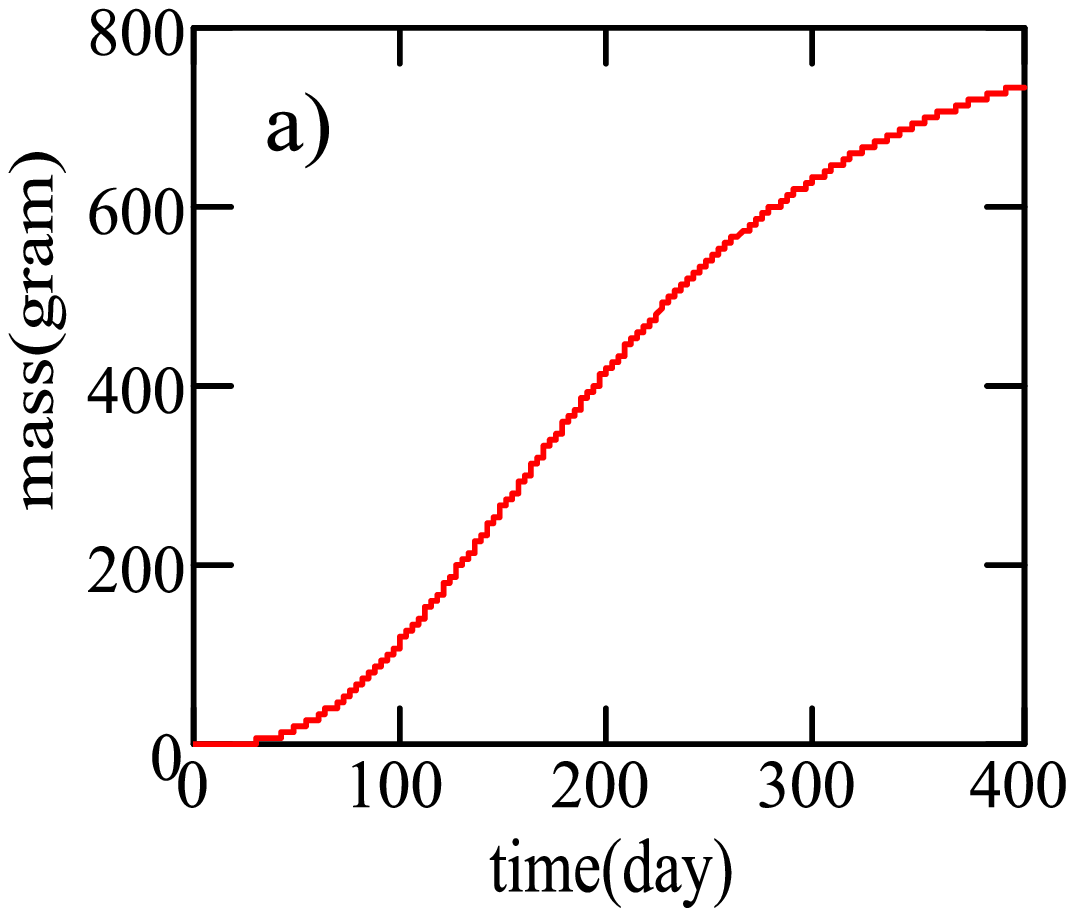}
 \end{minipage}
 \begin{minipage}[b]{0.47\textwidth}
 \centering
 \includegraphics[width=.6\textwidth]{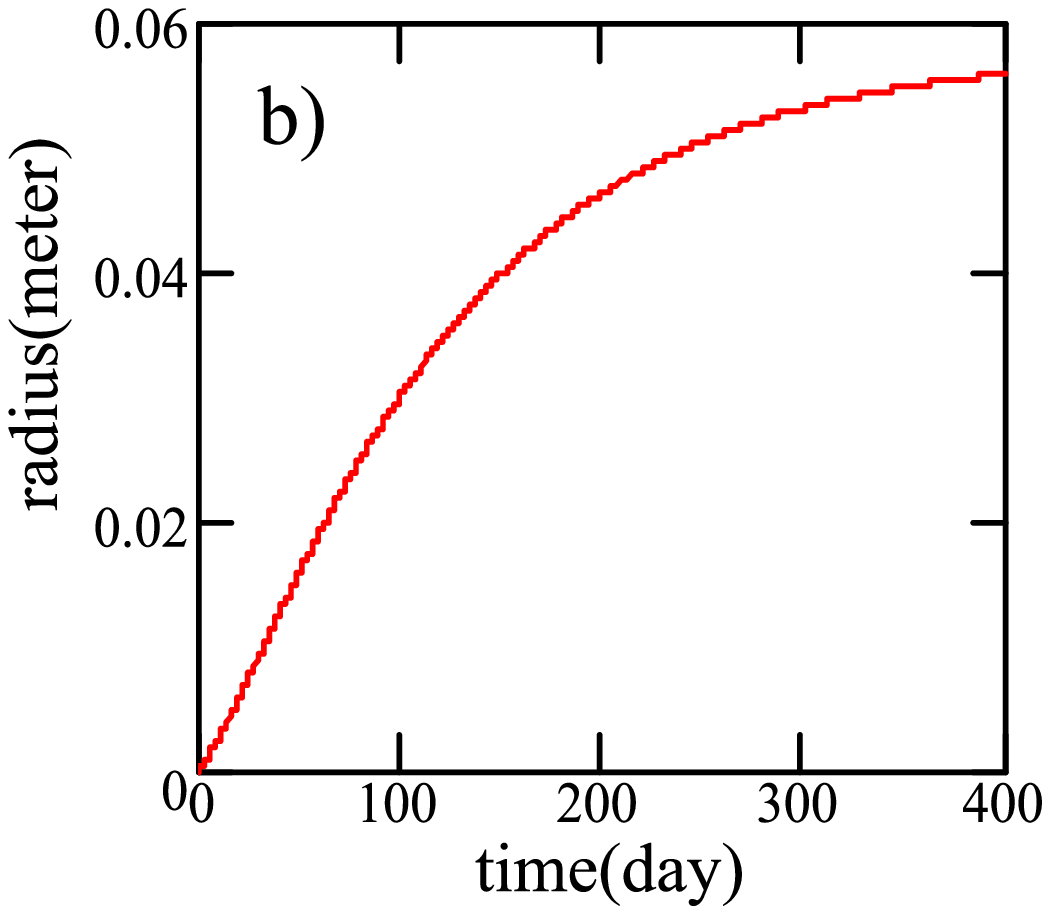}
 \end{minipage}
 \caption{Tumor mass evolution vs. time in $a)$ and tumor radius evolution vs. time in $b)$ given by the solution of the WBE equation, both graphs for $M_{\infty}\simeq 800g$ and $\tau_{1}\simeq 107$ $days$.}
 \end{figure}\label{m(t)r(t)}

\begin{figure}[h]
   \centering
   \includegraphics[width=.4\textwidth]{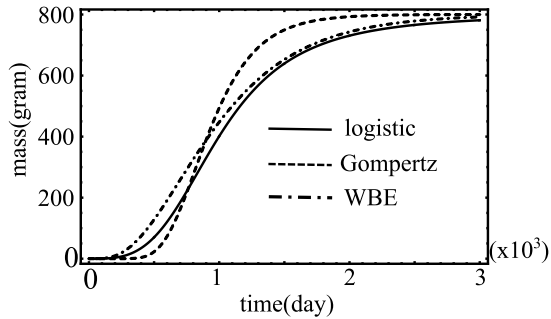}
   \caption{Comparison between WBE (solid), Gompertz (dot) and Logistic (dash) curves imposing for the $A(t)$ function the following expression: $(A(t)=(1+(t/\tau_3)^3)^{\alpha})$, with $\tau_3 \sim (10^{-3} \tau_1)$ and  $\alpha\simeq 1.13$; the all graphs are obtained starting from a very small initial cancer mass ($m(0)=3 \cdot 10^{-9} g$) reaching the final value equal to $M_{\infty}\simeq 800g$; the value of the $\tau_{1}$ parameter (the time necessary to reach $16\%$ of the final mass ) has been fixed to $500$ $days$.} \label{comparison}
  \end{figure}

The Gompertz curve is usually exploited to fit the evolution of tumor masses. In Fig. \ref{comparison} we have compared $WBE$ and Gompertz

\begin{equation}\label{Gompertz}
m(t)=M_{\infty}e^{ke^{-\frac{t}{\tau_2}}}, \qquad k=\ln\left(\dfrac{m_{0}}{M_{\infty}} \right)
\end{equation}
and found that, albeit similar the relevant predictions are not coincident (being $\tau_2 \sim 0.5 \tau_1$). In order  to exploit a further comparison tool, we have considered the Verhulst (logistic) function \cite{Tsoularis}
\begin{equation}\label{logistic}
m(t)=m_{0}\dfrac{A(t)}{1+\dfrac{m_{0}}{M_{\infty}}(A(t)-1)}
\end{equation}
where the function $A(t)$  is usually an exponential. Here, for further convenience, it will be assumed to be non-singular, at least once differentiable and such that $A(0)=1$, $\lim_{t\rightarrow\infty}A(t)\rightarrow\infty$.\\

The logistic growth is a paradigmatic tool in evolution problems and it can be exploited either to interpolate between WBE and Gompertz,  or as bench-mark with the other models.\\

\begin{figure}[h]
 \centering
 \includegraphics[width=.6\textwidth]{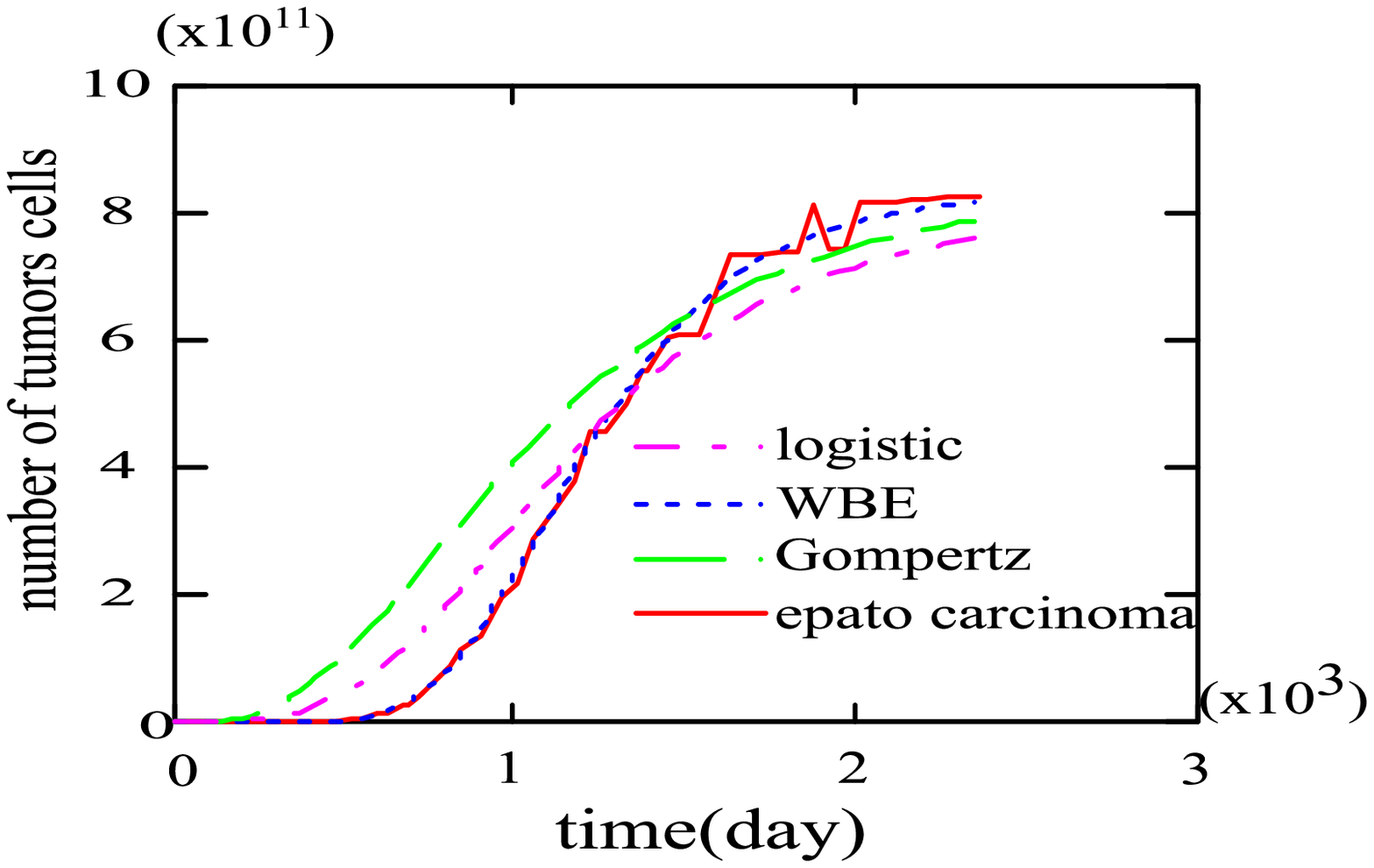}
 \caption{Comparison between $WBE$ growth curve and the code predictions of ref. $[12]$, for a hepato cellular-carcinoma.} \label{comparison2}
\end{figure}

 In Fig. \ref{comparison2} we make a comparison between the predictions of eq. \ref{WBE}
 %\ref{Gompertz}\ref{logistic}
   and those from the code of ref. \cite{Haustein}, which has been worked out using the assumption that a malignant cell inside a
  tumor has three possibilities: mitosis, apoptosis and migration into another department. Each one of the processes  follows an exponential distribution with a given characteristic time and occurs with restriction  of a non-overlapping chronology.\\

 The model, being based on purely probabilistic arguments, is a further independent test of the deterministic procedures we are dealing with and that will be further exploited in the forthcoming sections to treat the dissemination of metastases.\\

Let us now go back to the problem of angiogenesis which can be assumed to be induced by some chemical signals (transmitted by the malignant cells) which trigger the formation of vessels \cite{Savage}. They
% Fig. \ref{capNetGrowth},
start to branch in order to feed an increasing mass. The network or web strategy is dynamical and opportunistic, and for this reason is very hard to be modeled. Whatever complexity occurs, mass conservation cannot be escaped. Furthermore, according to Murray law, the branching strategy is such that after a certain number of division the capillaries combine to ensure the same nourishment for cancer cells inside any region of the tumor.\\

In a very first approximation we can just say that the temporal evolution of the number of capillaries follows the evolution of the tumor mass (or viceversa) and the characteristic times are the same.
However, the formation of a necrotic core in the center of cancers due to the hypoxic effect, that is associated with insufficient vascularization, cannot be arranged in this na\"{\i}ve description.
Rather, this process can be interpreted as the result of the cancer evolution strategy aimed at infiltrating the healthy tissue outside its outer surface and this seems to be in conflict with the previous discussion on Murray's law.
We can however assume that, despite the capillary system's intent to ensure uniform feeding, the restriction of the capillary radius, after successive branching, increases the hydraulic impedance of the vessel above the simple limit based on the assumption of a Poisseille fluid.
This mechanism leads to a significant increase of the power lost by dissipation, thus causing insufficient diffusion of nutrients towards the inner regions of a spheroidal cancer mass (see the concluding section).\\

Furthermore the dynamical behavior of the cancer mass and of the associated capillary network follows inverse patterns.  That is, the tumor mass grows outward further into the microenvironment while the capillary web diffuses towards the cancer cells. When the characteristic times of the two evolutions are well matched, nutrients  spread uniformly. With the increase of mass changes in density due to a variety of factors (overlapping between healthy and malignant tissues, different cell adhesion mechanisms, et al.)\cite{Wise} this may cause variations of the branching circuit, thus determining a non uniform distribution of nutrients and thus the emergence of hypoxic regions \cite{Cristini,Zheng}.\\

Even though the mathematical description of these processes has been brought to a high level of sophistication, we will make a comparison with the simpler procedure developed here by showing that the prediction relevant to macroscopic quantities such as the cancer radius (and thus the mass) evolution provide non conflicting results.\\

\section{Use of simple model for clinical purposes}
%\numberwithin{equation}{chapter}
%\markboth{\textsc{\chaptername~\thechapter. 2.	Use of simple model for clinical purposes}}{}

%In the previous section we have mentioned a few “popular” models used to study cellular growth, in general, and cancer evolution in particular. A systematic analysis to encompass all the various (unidimensional) models within a unified point of view has been undertaken in ref. \cite{Delsanto} and is reported below, because of its relevance to the topics we will treat in the following.\\

 Albeit not explicitly stated in the previous section, it is evident that we have assumed that cellular growth, being strictly dependent on the actual replicating population,  can be described with the greatest generality by equations of the type \cite{Delsanto}
\begin{equation}\label{dm}
\dfrac{dm}{dt}=a(t,m)m
\end{equation}
Where $a(t,m)$  is an explicit  function expressing the instantaneous growth rate. Its explicit form specifies the type of evolution model. Leaving for the moment such a function unspecified, we use the method put forward in \cite{Delsanto}, by assuming that its time derivative can be expanded as
\begin{equation}\label{seriepot}
\dot{a}=\alpha+\beta a+\gamma a^{2}+\dots
\end{equation}
Being $\alpha,\beta,\gamma$ constants to be determined which allows the evaluation of the different orders of solutions.\\

Since eq. \ref{seriepot} can be easily rescaled to nullify $\alpha$,  the following approximate solutions can be found:
\begin{enumerate}
\item[1)] when $\dot{a}=0$   eq. \ref{dm} will reproduce an exponential function, being the function  $a(t,m)$  a constant, which is therefore understood as the growth rate;
\item[2)] when $\dot{a}=\beta a$ eq. \ref{dm} can be reinterpreted as the Gompertz equation;
the instantaneous growth rate is indeed $a=a_{0}e^{\beta t}$ , which once inserted in \ref{dm} yields $m(t)=m_{0}e^{\frac{\alpha_{0}}{\beta}(e^{\beta t}-1)}$  recognized as the Gompertz function \ref{Gompertz}; provided that $\beta=-\dfrac{1}{2\tau_{1}}$, $a_{0}=\dfrac{1}{\tau_{1}}\ln\left(\dfrac{M_{\infty}}{m} \right) $  , the corresponding differential equation can accordingly be written as
\begin{equation}
\dfrac{dm}{dt}m=\dfrac{m}{2\tau_{1}}ln\left(\dfrac{M_{\infty}}{m} \right)
\end{equation}
\item[3)]	A similar approach employed when $\dot{a}=\beta a+\gamma a^{2}$ yields both logistic \footnote{According to eq. \ref{logistic} the solution of the logistic equation holds also if the function $A(t)$ is not an exponential.}  and $WBE$ model differential equations, respectively
\begin{equation}\begin{split}
& \dfrac{d}{dt}m=\dfrac{A^{'}(t)}{A(t)}\left(1-\dfrac{m}{M_{\infty}} \right)m,  \qquad A(t)=e^{-\beta t}\\
&  \dfrac{d}{dt}m=\dfrac{4}{\tau_{1}}\left[M_{\infty}^{\frac{1}{4}}m^{\frac{3}{4}}-m \right]
\end{split}\end{equation}
\end{enumerate}
It is evident that whatever model we use the associated growth equation can be cast in the form
\begin{equation}
\dfrac{d}{dt}m=mF(m)
\end{equation}

The consequences associated with different forms of the function $F(m)$ will be discussed in the concluding section.\\

Before going further we want to stress the consistency of this approach with other works in which the problem has been treated in very complete terms, either mathematically and numerical \cite{Cristini,Zheng}, and we compare the associated predictions relevant to the tumor radius evolution.\\

This is a quantity of noticeable importance for the analysis of the experimental data. If we assume the mass to be spherical in shape, with the evolution fixed by a Verhulst equation, we obtain for the radius the associated differential equation
\begin{equation}\label{Verhulst}
\dfrac{d}{dt}r=\dfrac{r}{3\tau_{1}}\left[1-\left(\dfrac{r}{R_{\infty}} \right)^{3}  \right]
\end{equation}
Or by \footnote{$R_{\infty}$, $\tau_{1,2}$, althought derived from the WBE model constant, can be treated as free parameters.}
\begin{equation}\begin{split}\label{Verhulst2}
& \dfrac{d}{dt}\tilde{r}=\dfrac{1}{\tau_{1}}\left(\tilde{r}^{\frac{1}{4}}-\tilde{r} \right) \\
& \tilde{r}=\dfrac{r}{R_{\infty}}
\end{split}\end{equation}
if the mass evolution is ruled by the WBE equation.\\

Both equations predict a bounded evolution, that occurs for sufficiently large times with respect to the characteristic time of the system.\\

The previous equations can also be interpreted as the velocity growth with respect to the radius itself. Previous researches have addressed the same problem in more general terms and in particular in ref. \cite{Zheng} this quantity has been expressed as
\begin{equation}\label{Verhulst3}
V=-\vec{n}\times (\vec{\nabla}p)_{\sum}+G\vec{n}\times (\vec{\nabla}\Gamma)_{\sum}-AG\dfrac{\vec{n}\times(\vec{x})_{\sum}}{3}
\end{equation}
Where $p$ and $\Gamma$ play the role of pressure and nutrient concentration, $\sum$ is the tumor surface (with a local curvature $\kappa$), $\vec{n}$ is its outward normal and $\vec{x}$ is the vector position in space. Furthermore $G$ and $A$
 are model parameters expressed in terms of the intrinsic time scales of the evolution associated with the cell mitosis and apoptosis rates.\\

 In the case of a radially symmetric tumor, the previous equation can be written (in our notation) as (see \cite{Zheng} for further comments)
 \begin{equation}\label{VerhulstSimm}
 \dfrac{d}{dt}\tilde{r}=-\dfrac{AG}{3}\tilde{r}+G\left(\dfrac{1}{\tanh (\tilde{r})}-\dfrac{1}{\tilde{r}} \right)
 \end{equation}
 Bounded growth in this model occurs only for $AG>0$ , G can be exploited to redefine the time. By expanding the $\tilde{r}$ dependent term in square brackets, we find
 \begin{equation}
 \dfrac{1}{\tanh (\tilde{r})}-\dfrac{1}{\tilde{r}}\backsimeq \dfrac{\tilde{r}}{3}-\dfrac{\tilde{r}^{3}}{45}+\dfrac{2\tilde{r}^{5}}{945}
 \end{equation}
 thus reducing eq. \ref{VerhulstSimm} to the simpler form
 \begin{equation}\label{Verhulst4}
\dfrac{d}{dt}\tilde{r}=G\left(1-\dfrac{A}{3}\right)\tilde{r}\left[1-\dfrac{\tilde{r}^{2}}{45\left(1-\frac{A}{3} \right) } \right]
 \end{equation}
 With suitable redefinition of the variables we find that the above equation can be rewritten as
 \begin{equation}\begin{split}\label{Verhulst5}
 & \dfrac{d}{dt}\tilde{r}=\dfrac{\tilde{r}}{\tau^{*}}\left[1-\left(\dfrac{\tilde{r}}{\tilde{r}_{\infty}} \right)^{2}  \right] \\
 & \dfrac{1}{\tau^{*}}=G\left(1-\dfrac{A}{3} \right), \qquad  \tilde{r}_{\infty}=\sqrt{45\left(1-\dfrac{A}{3} \right)}
\end{split} \end{equation}

\begin{figure}[h]
 \centering
 \includegraphics[width=.5\textwidth]{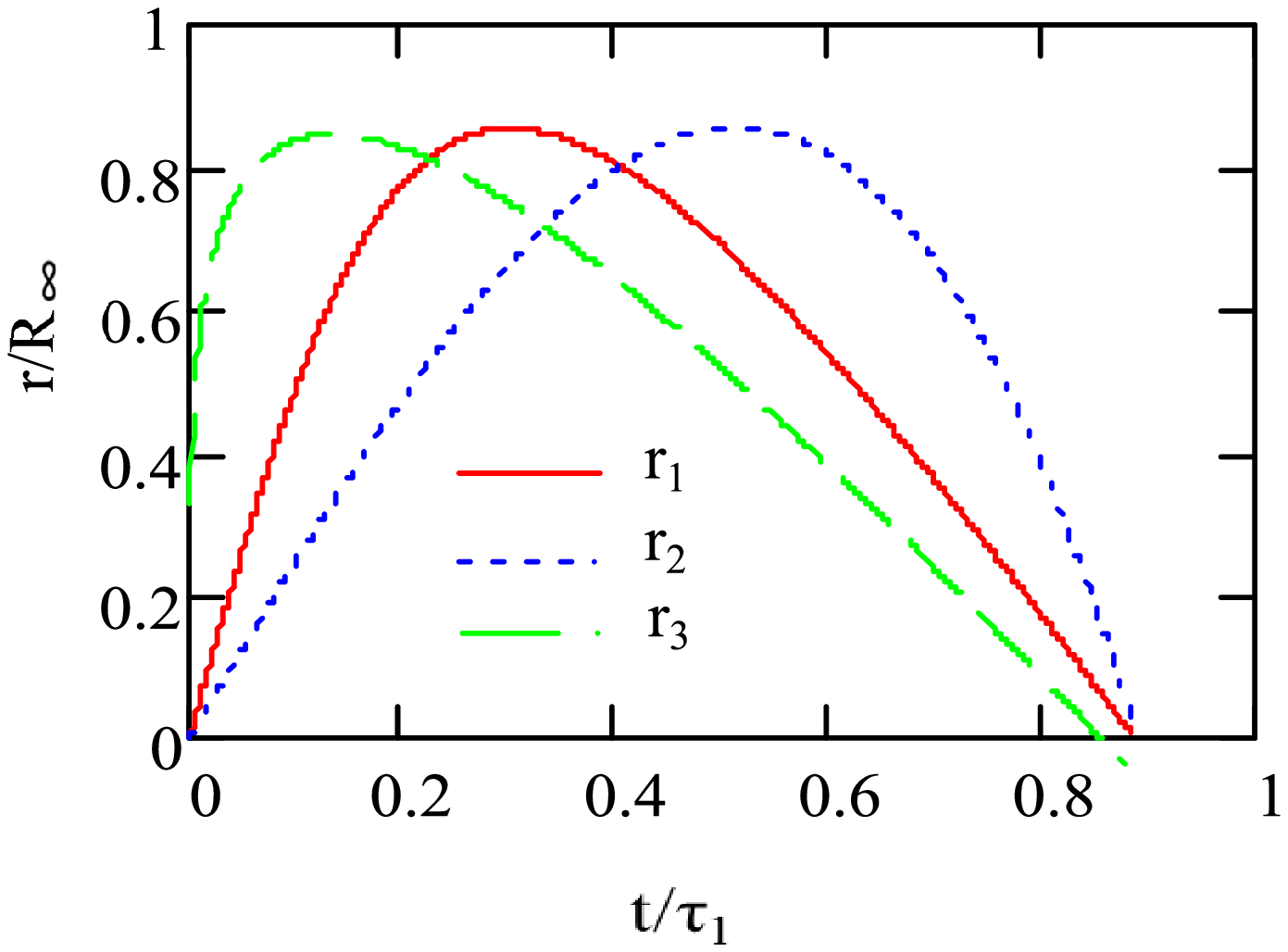}
 \caption{Expansion velocity vs. the radius predicted by eqs. \ref{Verhulst} (slash), \ref{Verhulst2} (dot), \ref{VerhulstSimm} (continuous).} \label{radiusVeloc}
\end{figure}

It is evident now that this last equation \ref{Verhulst5} belongs to the same families of those discussed so far. A comparison between eqs.  \ref{Verhulst} \ref{Verhulst2} \ref{VerhulstSimm} after a suitable rescaling of the variables is reported in Fig. \ref{radiusVeloc}, which evidently shows that qualitative similar behavior is obtained for the three cases.\\

In a forthcoming investigation we will provide a more detailed comparison by deriving the mass evolution equation from eq. \ref{VerhulstSimm}\\

We have continued to use the tumor radius evolution because it is an extremely useful quantity in clinical practice.\\

The use of such a pragmatic point of view suggested in the past in ref. \cite{Friberg} considers the “gross” growth rate of human malignancies, as studied in their hosts, for medical decision making.\\
Within such a context it has indeed been shown \cite{Michaelson} that mortality (i.e. the fraction of patients not surviving a given time period) and lymph node positivity (fraction of patients with at least one affected node) increase with ncreasing size of the primary tumor.\\
The problem has been thoroughly treated in the past (see refs. \cite{Michaelson,Leon}), where the analysis of data and associated statistical considerations have allowed the following quantitative link between mortality and tumor size
\begin{equation}
L(r)=1-e^{-qr^{z}}
\end{equation}

\begin{figure}[h]
 \centering
 \includegraphics[width=.5\textwidth]{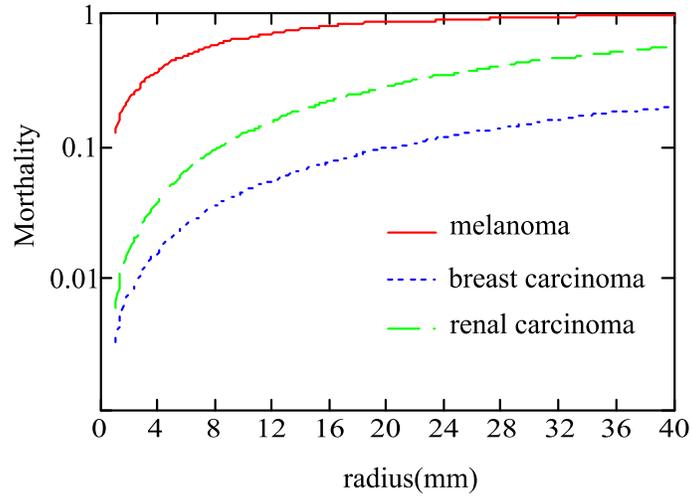}
 \caption{Mortality vs. radius (mm) for three types of tumors,
melanoma ($q=0.134$ ,$z=0.89$ continuous line), breast carcinoma ($q=0.0061$,$z=1.33$  dash line), renal carcinoma ($q=0.0033$,$z=1.14$ dot line).
 } \label{lethality}
\end{figure}

Where $q,z$  depend on the type of tumor as also illustrated in Fig. \ref{lethality}, where we have reported the mortality (after $15$ years) for three different types of tumors (melanoma, breast carcinoma, renal cell carcinoma)\\

It is now possible to combine the data from the growth analysis to get an idea of the mortality  along the time history of the tumor itself.\\

The use of the $WBE$ parameterization for the mass radius evolution (eq. \ref{raggio}) yields for the mortality the following formula

\begin{equation}\begin{split}
& L_{WBE}=1-P_{\infty}^{\left[1-\left(1-\sqrt[4]{\frac{m_{0}}{M_{\infty}}} \right)e^{-\frac{t}{\tau_{1}}}  \right]^{\frac{4}{3}z} }\\
& P_{\infty}=e^{-qR_{\infty}^{z}}
\end{split}\end{equation}
which expresses the mortality at a certain time of the cancer evolution.\\

\begin{figure}[h]
 \centering
 \includegraphics[width=.5\textwidth]{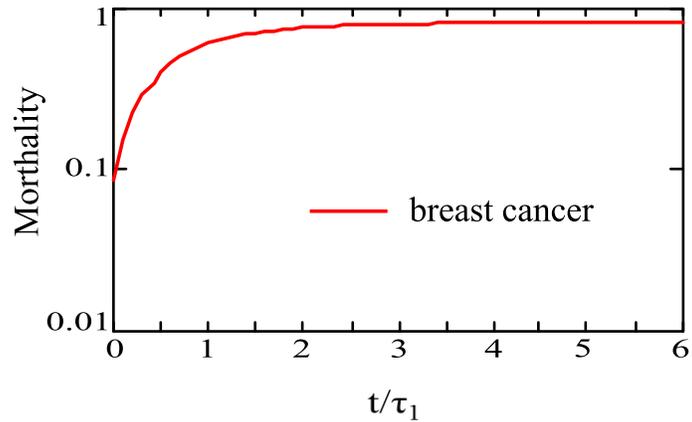}
 \caption{Evolution vs time $ \left(\tau=\frac{t}{\tau_{1}} \right)$ of breast cancer mortality $\left(M_{\infty=800g} \right) $.
 } \label{breastCancer}
\end{figure}

In Fig. \ref{breastCancer} we have plotted the mortality and the radius vs. time for the breast cancer .\\

A way to exploit the above formulae from the clinical point of view could be to monitor the cancer data of a patient, “predict” the relevant tumor development, infer the associated mortality and, finally, deduce the most appropriate strategy of treatment. \\

In the forthcoming section we will further implement the model by including the effects associated with the spatial spatial expansion of the tumor cells inside the host organ.

\section{Metastatic Tumor Spreading and Fisher Equation}
%\numberwithin{equation}{chapter}
%\markboth{\textsc{\chaptername~\thechapter. Tumor spreading and Fisher Equation}}{}

Although the previous simple models have some “practical” outcome, they do not contain any information on how the tumor spreads inside its own spatial environment. To understand how the combined mechanism of spatial diffusion and cellular growth may occur, we use a reaction diffusion equation of the Fisher type \cite{Fisher}, which combines spatial diffusion regulated by an ordinary heat equation and cell growth models of the previously discussed form. The equation we will deal with, known as a reaction diffusion problem, is specified by the following non-linear PDE
\begin{equation}\begin{split}\label{reactDiff}
& \partial_{t}N=\left[D_{x}\partial_{x}^{2}+D_{y}\partial_{y}^{2}+D_{z}\partial_{z}^{2} \right]N+NF(N) \\
& N\mid_{t=0}=N_{0}f(x,y,z)
\end{split}\end{equation}

Where we have denoted by $N$ the cell density as a function of the three spatial variables, and $D$ the diffusion constants, which will be assumed to be constant in time and space. The non-linear  part should be simply reinterpreted and indeed $M_{\infty}$  should be replaced by $N_{\infty}$ , namely the asymptotic cell density, which is linked to the  asymptotic mass by $M_{\infty}=m_{c}N_{\infty}V_{\infty}$.\\

We will perform a preliminary analysis of the problem by assuming that eq. \ref{reactDiff} be a "canonical" Fisher equation with $F(N)=\lambda\left(1-\dfrac{N}{N_{\infty}} \right) $.
In handling with this non-linear problem we should have information about its free constants  (diffusion and growth coefficients, asymptotic density and volume),
 they will then be exploited as input parameters for the numerical integration. Some preliminary indications on how to proceed are given below. \\

\begin{figure} %[h]
 \begin{minipage}[b]{0.47\textwidth}
 \centering
 \includegraphics[width=.9\textwidth]{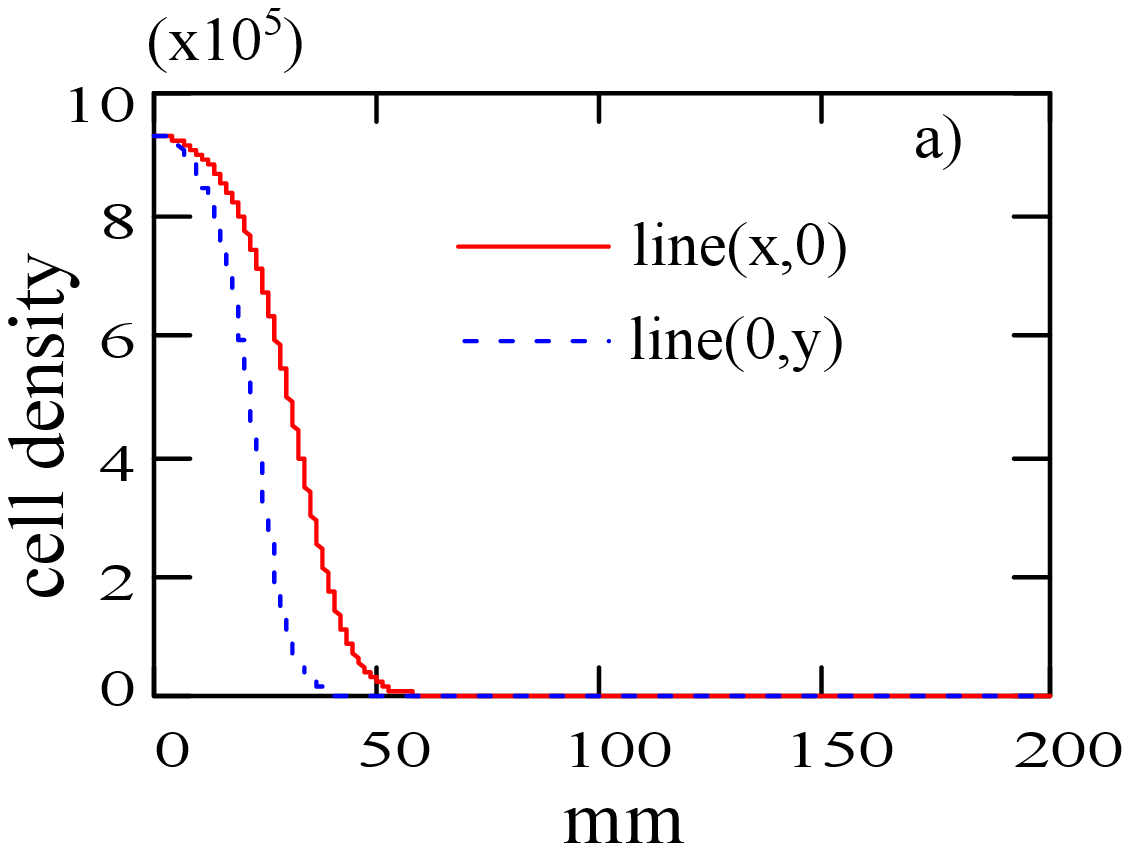}
 \end{minipage}
 \begin{minipage}[b]{0.47\textwidth}
 \centering
 \includegraphics[width=.9\textwidth]{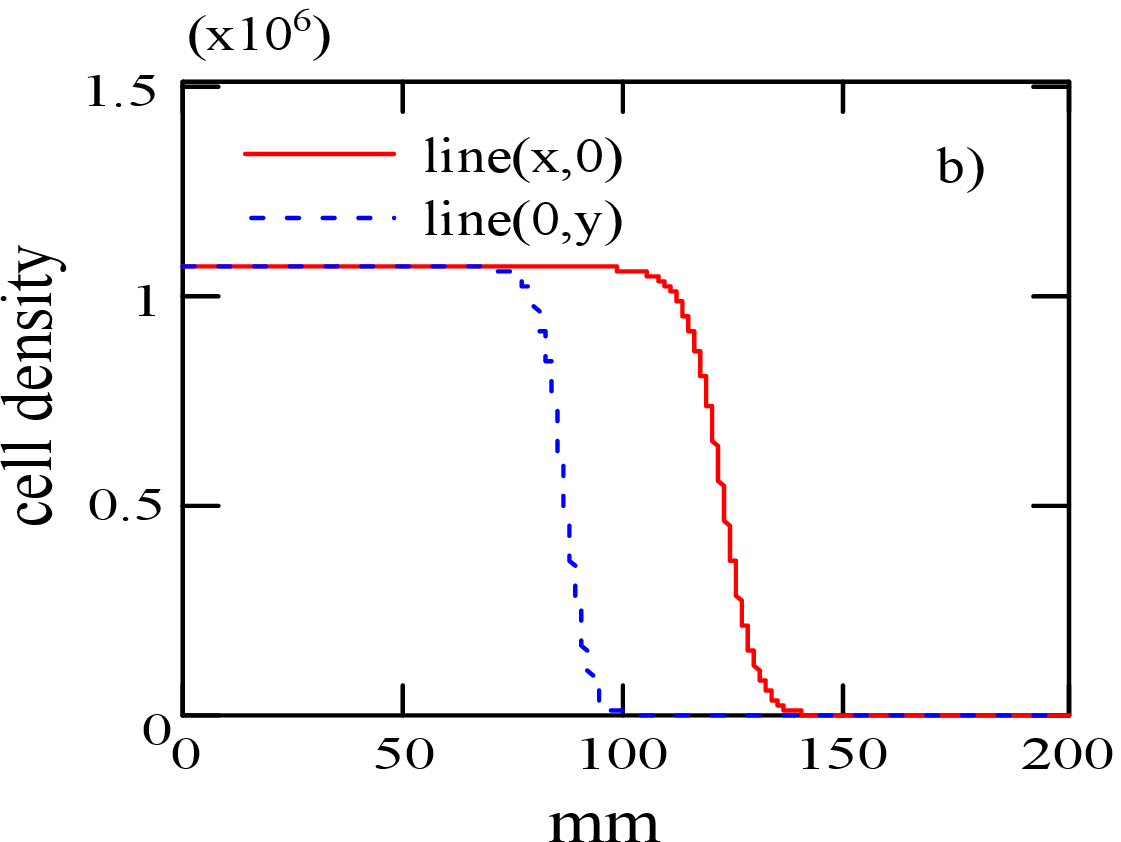}
 \end{minipage}
 \caption{Cell density evolution in a two dimensional domain with different diffusion coefficients in each direction: $D_{x}\left[\frac{mm^{2}}{day} \right]=0.2 $,$D_{y}\left[\frac{mm^{2}}{day} \right]=0.1$; these graphs illustrate the density trend along two specific directions, for fixed y (continuous line,y=0) and for fixed x (dashed line, x=0), at two different times: a) 270 days and b) 570 days.} \label{lethality1}
 \end{figure}

\begin{figure}[h]
   \centering
   \includegraphics[width=.5\textwidth]{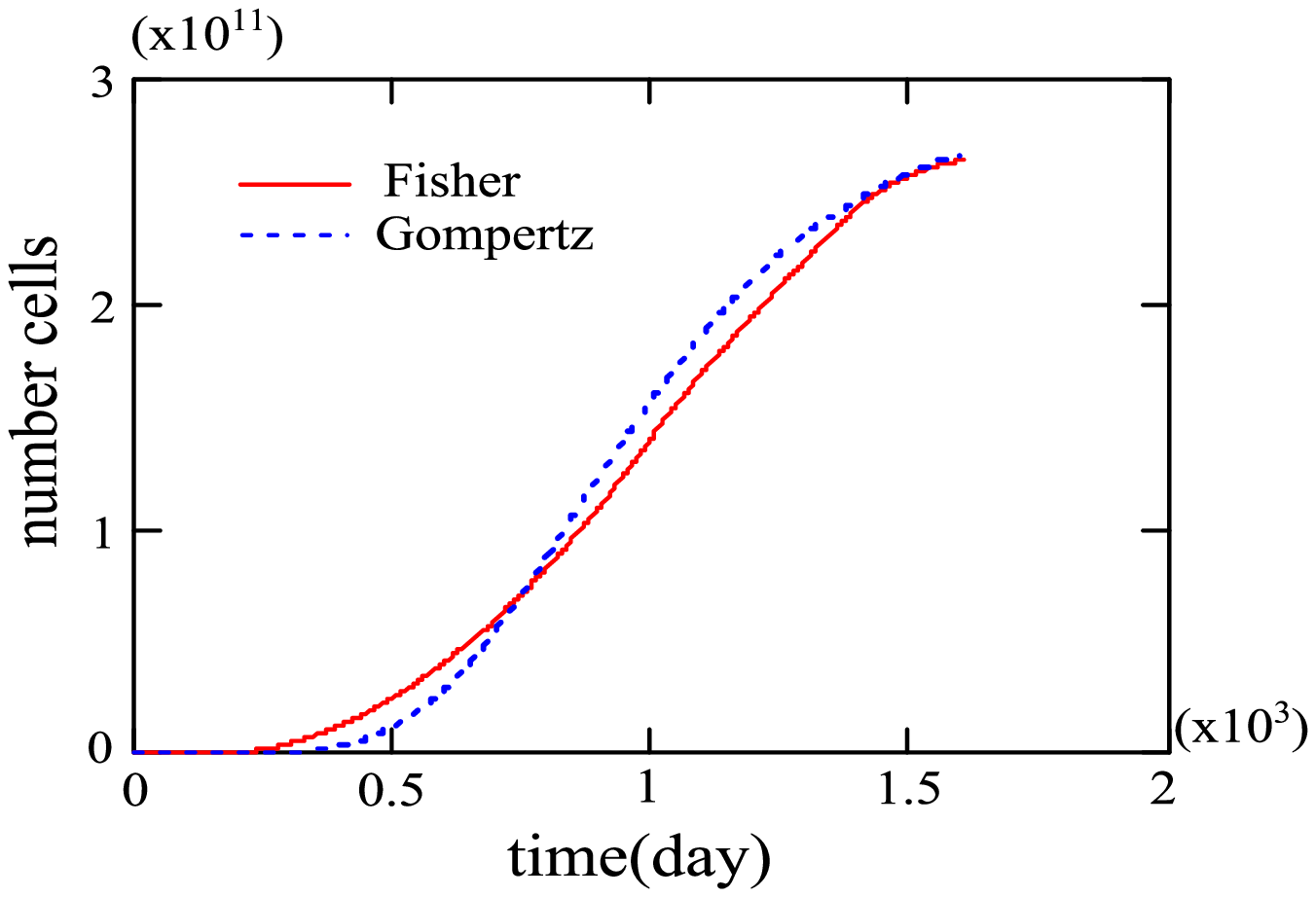}
   \caption{Number cell evolution vs. time (days) for the two dimensional Fisher equation (continuous line), and Gompertz equation
   $A(t)=e^{rt}$, $r\left[day^{-1} \right]=0.05 $ , $D_{x}=D_{y}=D \left[ \frac{mm^{2}}{day}\right]=0.15 $ ,  $N_{\infty}=7\cdot 10^{6}$ , $\tau_{1}\simeq 600$.
   } \label{numbercells}
  \end{figure}

The diffusion parameters $D$ have the dimension of a surface divided by a time. They have been extensively
 studied in the literature (see e.g. refs. \cite{Ferreira}) and they depend on the type of tissue the tumor is spreading in: here we will use an average value taken from the available data
 $D\simeq (0.1-0.2)\dfrac{mm^{2}}{day}$  , while the growth rate $\lambda$   is taken to be on the order of
 $\lambda\simeq (0.01-0.05)\dfrac{1}{day}$ .
  Finally, to estimate $N_{0,\infty}$ we need some further remarks: according to the discussion of the introductory section we can conclude that the initial number of cells per unit volume is $N_{0}\simeq 10^{4}\dfrac{cells}{mm^{3}}$; regarding the threshold density we can use e.g. the predictions of the WBE or Gompertz equation to infer that for the cases considered in Figs. \ref{lethality1} and \ref{numbercells} the asymptotic number of cells is larger than $10^{11}$  over a volume (supposed to be spherical) on the order of $V_{\infty}\simeq 10^{5}mm^{3}$ , calculated for an asymptotic mass of hundreds of grams.\\

The Fisher equation \ref{reactDiff} has been solved using a full numerical procedure employing either the finite difference method \cite{Blanes} or the evolution operator technique, based on the exponential symmetric split procedure \cite{Dattoli3}. The two methods, used as reciprocal benchmarks, have been further corroborated by means of a semi-analytical solution, which was derived by merging the solutions of the Verhulst and of the heat equation separately.\\

In ref. \cite{Dattoli4} it has been shown  that a very accurate, albeit simple, approximate solution of eq. \ref{reactDiff} is provided by the following expression

\begin{equation}\begin{split}\label{solreactDiff}
& N(x,y,z)=N_{0}\dfrac{\Phi (x,y,z)}{1+\dfrac{N_{0}}{N_{\infty}}\left[\Phi (x,y,z,t)-\Phi (x,y,z,o) \right] } \\
& \Phi (x,y,z,t) = A(t)\dfrac{e^{-\left[\left(\dfrac{x}{\sum _{x}(t)\sigma_{x}} \right)^{2}+\left(\dfrac{y}{\sum _{y}(t)\sigma_{y}} \right)^{2}+\left(\dfrac{z}{\sum _{z}(t)\sigma_{z}} \right)^{2}  \right] }}{\sum _{x}(t)\sum _{y}(t)\sum _{z}(t)}\\
& \sum _{k}(t)=\sqrt{1+4\dfrac{D_{k}t}{\sigma _{k}^{2}}}, \qquad k=x,y,z\\
& N_{0}=\dfrac{n_{0}}{\sigma_{x}\sigma_{y}\sigma_{z}}
\end{split}\end{equation}
Which assumes that the initial distribution is provided by the product of three Gaussian functions.\\

The solution of the Fisher equation “estimated” in eq. \ref{solreactDiff} captures the essential features of the phenomenology we are interested in; it contains indeed the “logistic” behavior including the saturation and the spreading of the diffusion, which follows a standard heat behavior. In Fig. \ref{lethality1} we have reported the spatial growth as predicted by eq.
 \ref{solreactDiff}, which is characterized by a kind of “rectangularisation” of the initial Gaussian
  distribution, hence in a more technical language we can say that the initial Gaussian tends to become a super Gaussian\footnote{A super Gaussian is specified by the function $e^{-\mid x\mid ^{m}}$ where the index m is any real. In this specific case it is an increasing function of time and depends on the wave front velocity.}.\\
%  \footnote{A super Gaussian is specified by the
%   function $e^{-\mid x\mid}^{m}$, $m>1$  ,
%    the index $m$ is any real.
%     In this specific case it is an increasing function
%      of time and depends on the wave front velocity
%}
%\begin{figure} %[h]
%\begin{minipage}[b]{0.47\textwidth}
% \centering
%%\begin{tabular}{cc}
% \includegraphics[width=.8\textwidth]{Celldensity2.pdf}
% \caption{\label{Celldensity2} Cell density (number of cells per square $mm$, $N(x,o,t)$ continuous line, $N(0,y,t)$   dot line) Vs. spatial coordinate $(mm)$, at different times}
%\end{minipage}
%\hfill
%\begin{minipage}[b]{0.47\textwidth}
% %\end{figure}
%%\begin{figure}[h]
% \centering
% \includegraphics[width=.8\textwidth]{Celldensity3.pdf}
% \caption{\label{Celldensity3} $a)$ $ 270$ $days$, $b)$ $ 570$ $ days$ for $D_{x}\left[\dfrac{mm^{2}}{day} \right]=0.2 $ , $D_{y}\left[\dfrac{mm^{2}}{day} \right]=0.1$}
%\end{minipage}
% \end{figure}

% \begin{figure}[h]
%  %\centering
%  %\alignment
% % \subfloat[][mini-didascalia]{\includegraphics[width=.3\textwidth]{Celldensity1}} \hfill
% \subfloat[][mini-didascalia]{\includegraphics[width=.3\textwidth]{Celldensity2}} \\
% \subfloat[][mini-didascalia]{\includegraphics[width=.3\textwidth]{Celldensity3}}
%  %\includegraphics[width=.6\textwidth]{angiogen.png}
%  \caption{\textbf{Cell density (number of cells per square $mm$, $N(x,o,t)$ continuous line, $N(0,y,t)$   dot line) Vs. spatial coordinate $(mm)$, at different times $(170,270,570 days)$
%   for $D_{x}\left[\dfrac{mm^{2}}{day} \right])=0.2 $,$D_{y}\left[\dfrac{mm^{2}}{day} \right])=0.1$
%  }} \label{miniDidasc}
% \end{figure}

 In Fig. \ref{numbercells} we have reported the cell number evolution $\left(n(t)=\int N(x,y,t)dS\; \right) $ for the two dimensional case, along with the comparison of the $1$
 dimensional model (Gompertz) predictions \footnote{The comparison with the Gompertz equation has been done by considering the following relationship between exponential
 growth rate and characteristic time $\tau_{1}\simeq \ln\left(\frac{M_{\infty}}{m_{0}} \right) $}.\\

\begin{figure}[h]
    \centering
    \includegraphics[width=.6\textwidth]{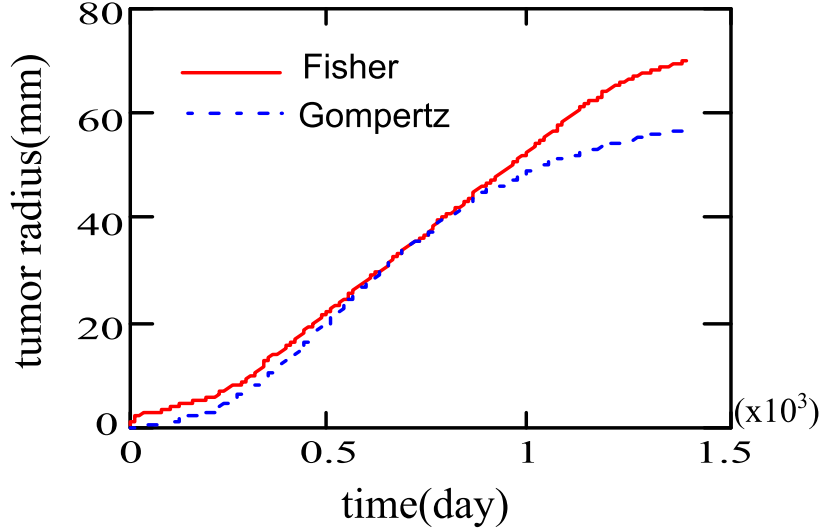}
    \caption{Tumor mass radius (mm), same parameters as in Fig. \ref{lethality1}, vs time comparison Fisher (continuous line) and Gompertz (dot line) eqs.} \label{radius}
   \end{figure}

 The toy model solution of the Fisher equation and the $1D$ case are reported in Fig. \ref{numbercells} and the agreement is good. In Fig. \ref{radius} we have reported the evolution of the tumor radius defined as $R(t)=\dfrac{\sqrt{\left\langle x^{2}\right\rangle }}{2}=\dfrac{\sqrt{\left\langle y^{2}\right\rangle }}{2}$   along with the same prediction from the Gompertz curve.\\

The one dimensional model and the use of the Fisher equation yield almost similar results, for macroscopic quantities like mass and radius, which should be benchmarked with a fully numerical computation.
Before addressing this aspect of the problem, let us stress a further tool which is getting increasing attention in the theory of tumor growth and diffusion, namely the use of PDE involving fractional derivatives, which are becoming increasingly popular within the context of anomalous diffusion \cite{HanertEtAl}.\\

The evolution equation which can be considered to deal with tumor diffusion is the same as eq. \ref{reactDiff} with the replacement
\begin{equation}\begin{split}
&\sum_{n=1}^{3}D_{n}\partial_{x_{n}}^{2}\rightarrow \sum_{n=1}^{3}\tilde{D}_{n}\partial_{x_{n}}^{2\alpha}, \qquad 0<\alpha<1 \\
& x_{1}=x,\; x_{2}=y,\; x_{3}=z
\end{split}\end{equation}
The diffusion coefficient, now denoted with a tilde, are measured in terms of $\left[\dfrac{mm^{\alpha}}{day} \right] $ . The toy model solution can be written as
%(see Appendix $A$)
\begin{equation}\begin{split}
& N(x,y,z,t)=N_{0}\dfrac{\Phi (x,y,z,t)}{1+\dfrac{N_{0}}{N_{\infty}}\left[\Phi (x,y,z,t)-\Phi (x,y,z,o) \right] } \\
& \Phi (x,y,z,t) = \dfrac{A(t)}{(2\pi)^{\frac{3}{2}}}\prod_{n=1}^{3}\sigma_{n}\int_{-\infty}^{\infty}e^{(ik_{n})^{2\alpha}\tilde{D}_{n}t+ik_{n}x_{n}}e^{-\left(\frac{\sigma_{n}k_{n}}{2} \right)^{2} }dk_{n}
\end{split}\end{equation}

The shape behavior of the distribution vs. time is given in Fig. \ref{radius}, where we have made the comparison with the prediction from the fractional and ordinary Fisher equation. The effect of the rectangularization tends to disappear and the initial Gaussian evolves into a distribution
 characterized by a longer tail. This is a signature of the anomalous diffusion and would suggest that if spatial tumor spreading is driven by anomalous diffusion, spatial spreading might be more invasive.\\

 This is only a qualitative remark and the problem will be treated with the necessary details in a forthcoming paper.
  As a preliminary remark we note that, within such a context, anomalous diffusion can be the result of the interaction between the tumor and the host tissue itself, i.e. its rigidity, haptotactic permission, its vascular architecture or lymph system, et al. \cite{Alberts}. As such, this is a composite of intrinsic tumor behavior and environmental
  ‘soil’ and we will see indeed that there are good reasons to believe that the coefficient $\alpha$ is linked to the fractal dimension of the tumor vascularization \cite{Maiti}\footnote{The fractional coefficient specifies the homogeneity of the tumor vasculature; tumors with homogeneous vasculature (e.g. hepatocellular carcinoma) exhibit a value of $\alpha =\dfrac{2}{3}$ as almost all cells on the surface have the same chance of metastasis. In the case of breast carcinoma with $\alpha=\dfrac{1}{3}$ only cells in special regions (on a line) metastasize.
For further comments see ref \cite{Maiti}.}. \\

  In the forthcoming section we will integrate the considerations developed so far with the mechanisms associated with metastatic dissemination.
\begin{figure}[h]
    \centering
    \includegraphics[width=.4\textwidth]{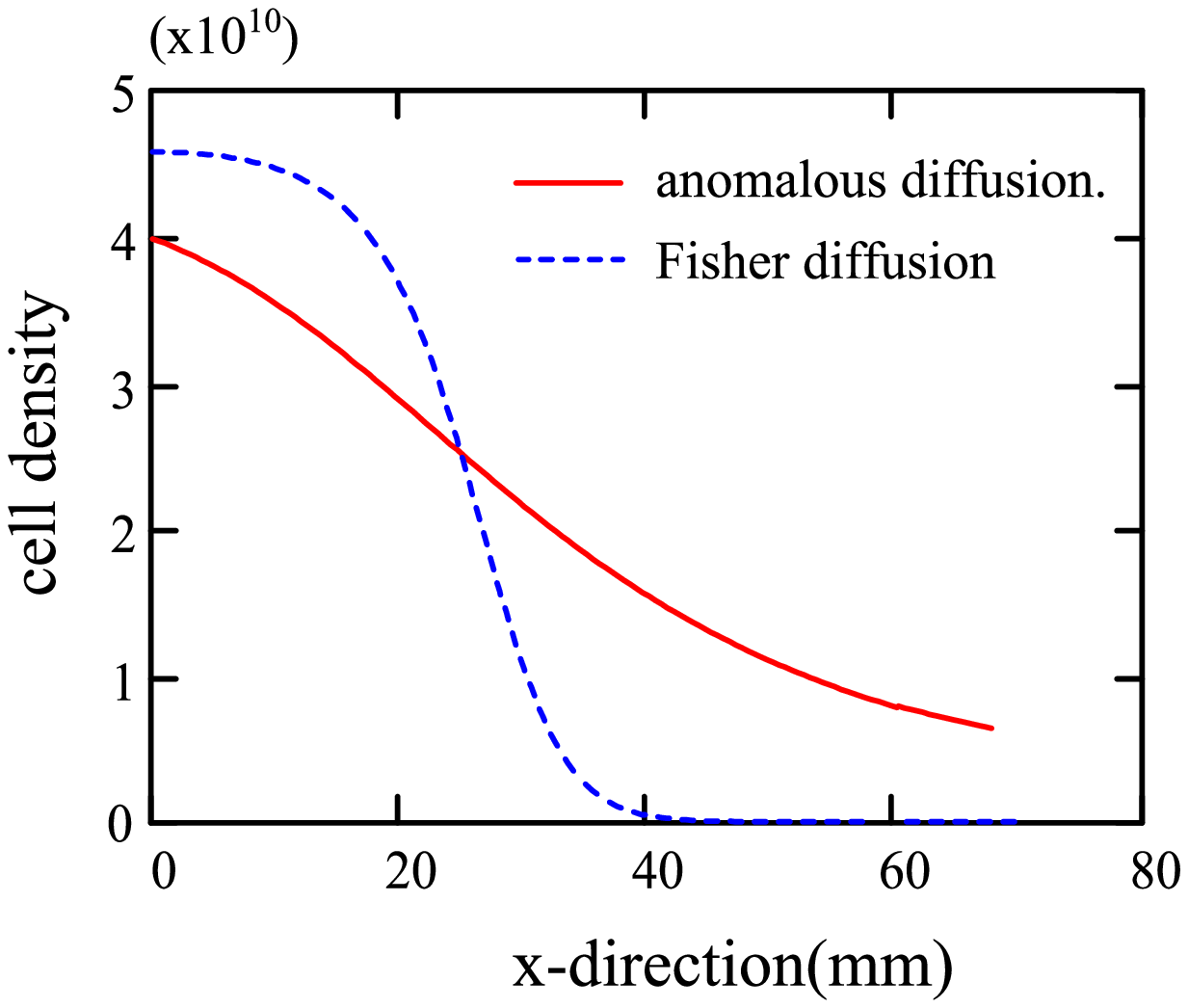}
    \caption{Cell density (number of cells per square $mm$) $N(x,0,t)$,  normal diffusion (dot line), anomalous diffusion (continuous line) vs. spatial coordinate $(mm)$, $t=500$ $days$ for $D_{x}\left[\frac{mm^{2}}{day} \right]=0.2$, $D_{y}\left[\frac{mm^{2}}{day} \right]=0.1$, $D_{x}\left[\frac{mm^{\alpha}}{day} \right]=0.2$, $D_{y}\left[\frac{mm^{\alpha}}{day} \right]=0.1$, $\alpha =\frac{3}{4}$.} %\label{cellDensity}
   \end{figure}\\\\

\section{Models of metastatic diffusion}
%\numberwithin{equation}{chapter}
%\markboth{\textsc{\chaptername~\thechapter. Models of metastatic diffusion}}{}

In section $2$ we have used a quantitative point of view linking mortality to the size of tumors. Mortality\footnote{We want to quote a curious identity emerging from statistical speculations \cite{Michaelson} concerning the link between the  tumor lethal mass and body mass $m$ in mammals, $m_{L}\backsimeq 0.05 \times m^{\frac{3}{4}}$.},
 although a well-defined statistical concept, does not clarify causality.
  Mortality often confused with the aggressiveness of the tumor itself can be assumed to be a consequence of the probability of metastasis. Such a probability cannot, however, be associated only with the tumor mass, it is indeed necessary to specify the metastatic potential characterizing a specific tumor (through e.g. empirical coefficients like the q-parameter reported in Fig. \ref{lethality}). Only after this distinction we can invoke the previously reported statistical evidence, to state that the probability of spreading metastasis increases with the mass of the primary tumor.\\

The metastasic process is the spread of cancer to another part of the body and more than $90\%$  of cancer associated death is due to metastasis.
 The pathway to such a diffusion process can be worded as: some tumor cells migrate, for some not yet specified reasons (in ref. \cite{Dattoli2,Deisboeck} a kind of energetic crisis is assumed, but this assumption alone is not sufficient), from the primary tumor mass
  into a neighbor blood vessel from which  they can colonize other organs, after moving outside the blood stream. They will be distributed in size on the basis of temporal and biological factors and are composed of only certain clones of the primary tumor constituents \cite{Iwata}.\\

 The link between the primary tumor dimension and the probability of metastatic spreading has been stressed in refs. \cite{Iwata,Dattoli2}. The problem has been treated using  genuine probability models: for instance in ref. \cite{Dattoli2} a Poisson distribution has been used to determine the probability of establishment of metastatic colonies with the increasing mass of the primary tumors, while in ref. \cite{Iwata} a kind of Fokker-Planck equation yielded the evolution of the densities of metastatic colonies. \\

The rate of metastases production is assumed to be provided  by \cite{Iwata}\\

\begin{equation}
\chi (Y,t)\simeq fY^{\alpha}
\end{equation}

Where $f$  is a constant with the dimension of the inverse of time, $Y$ is some characteristic quantity characterizing the size of the primary tumor (volume, number of cells…) and $\alpha$ is the fractal dimension of the relevant vasculature \cite{Savage,Maiti}.
 This last quantity is a characteristic of the tumor species and it can in principle be associated with the fractional coefficient of the modified Fisher equation, discussed in the previous section.\\

 A given metastatic cell can invade directly the vital organs, or migrate into a lymph-node. Each such metastatic cell promotes the formation of new tumors, or can even self-seed the primary tumor itself.\\

The natural questions arising after this discussion are therefore\\

$a)$	 How many secondary tumors do exist? \\

$b)$	Can they be considered separate entities or are they the components of a coherent biological complex delocalized in space? \\

Although at the moment any comment to point b) is just a matter of speculations, the possibility of a “coherent body” is not simply a metaphor and indeed there is experimental evidence of enhancement of the activity of dormant secondary tumors when the primary has been surgically removed or treated (chemically or with radiotherapy) \cite{Hanin}. \\

However, the question raised in a) is not far from our theoretical understanding and we can try providing an answer by using the model proposed in ref. \cite{Iwata,Stein}. The analysis put forward there allows following the evolution of the distribution function $\rho (Y,t)$, which represents the density of metastatic tumors (supposed to be far enough  to be considered independent entities) with a number of cells equal or larger than $Y$. The relevant equations are given below
\begin{equation}\begin{split}\label{iwata}
& \partial_{t}\rho(Y,t)=-\partial_{Y}\left[g(Y)\rho(Y,t) \right]  \\
& \rho(Y,0)=\rho_{0}(Y)\\
& g(1)\rho(1,t)=\int_{1}^{n(t)}\chi(\xi)\rho(\xi,t)d\xi +\chi(n(t))
\end{split}\end{equation}

with $g(Y)$  being the growth rate of the primary tumor.\\

From the mathematical point of view the first two lines in eq. \ref{iwata} represent a (first order) Cauchy  problem, which admits a unique solution, depending on the initial distribution  $\rho_{0}(Y)$ while the third line provides a kind of boundary conditions.\\

Since the associated mathematical problem has been treated in depth in a series of papers [see \cite{Stein} and references therein], here we will use an heuristic approach by pointing out that  the search of a solution is made difficult by the necessity of satisfying, consistently, the initial distribution $(\rho_{0}(Y))$ and the boundary conditions.
 We solve therefore the first order evolution equation, by means of the algebraic evolution method \cite{Dattoli3}. To avoid inconsistencies, we proceed by first solving the problem without assuming any specific form of the initial distribution, which is then  specified through the boundary condition.\\

In the case of a constant growth rate $\left(g(Y)=\dfrac{1}{\tau}Y\right) $, we obtain the following solution using the algebraic operator evolution method
%(see App. $B$ for the algebraic details)
\begin{equation}
 \rho(Y,t)=e^{-\frac{t}{\tau}\partial_{Y}Y}\rho_{0}(Y)=e^{-\frac{t}{\tau}}\rho_{0}\left(e^{-\frac{t}{\tau}}Y \right)
\end{equation}
The form of the initial distribution function can be determined from the boundary condition with $n(t)=e^{\frac{t}{\tau}}$ so that we find
 \begin{equation}
 \rho(Y,t)=f\tau e^{\left(\frac{\alpha}{\tau}+f \right)t }\dfrac{1}{Y^{\alpha+f\tau+1}}
 \end{equation}
Which yields that, at time $t$, the number of secondary tumors with a volume larger than $Y$, is
\begin{equation}\begin{split}
& N_{S}(t)\simeq\int_{Y}^{\infty}\rho(Y^{'},t)dY^{'}=\dfrac{f\tau}{\alpha+f\tau}e^{(\alpha+f\tau)\frac{1}{\tau}}\dfrac{1}{Y^{\alpha+f\tau}}\\
& Y<Y_{1}
\end{split}\end{equation}
Where $Y_{1}$  is the number of cells of the primary tumor.\\
\begin{figure}[h]
    \centering
    \includegraphics[width=.4\textwidth]{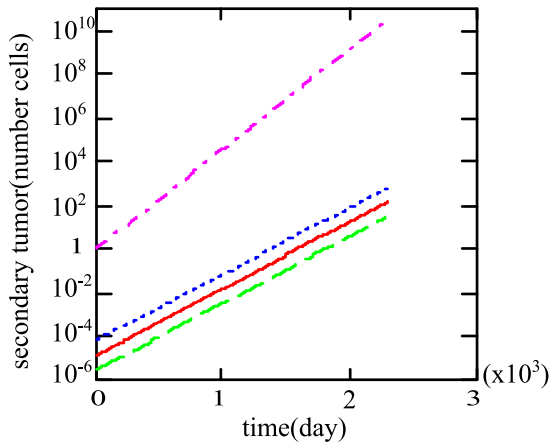}
    \caption{Evolution of the primary (dash-dot) and secondary tumors. Number of tumors with $10^{2}$ (dot), $10^{3}$  (continuous), $10^{4}$  (dash) cells. Total number of colonies with a number of cells up to $10^{6}$.} \label{evolPrSec}
   \end{figure}

Using for $f$  the currently quoted value \cite{Stein} $5\cdot 10^{-7}\left[cell\cdot day^{-1} \right] $ and  $\tau[day]=70$, $\alpha=\frac{2}{3}$, $t[day]=1000$, we obtain what is reported in Fig. \ref{evolPrSec}.\\

A tumor of monoclonal origin may have disseminated, when at the mass stage of $300g$, colonies of metastases which contain $10^{5}-10^{6}$ $cells$  , corresponding to secondary tumors of tens of grams.\\
This result is hampered by the fact that we have limited our analysis to that of linear growth and we did not include any spatial effect; this notwithstanding, our rather naive approach yields an idea of the temporal evolution of metastasis.

\section{Concluding Comments}

Cancer is a complex non-linear system with many interacting elements. Similar to other complex systems in nature, its organizational structure produces power laws and as such, all the relevant phenomenology should be framed within the context of the so-called self-organized-critical systems (SOC) \cite{Ascwanden}. The celebrated Kleiber law can therefore be a manifestation of the tumor as a SOC.\\

In the previous sections we have developed several analytical tools that model diverse aspects of the cancer physiology. We have used the Murray law to get a glimpse into the development of the relevant capillary network and to derive the relationship between the successive branching steps.
Althought the problem has been treated in previous works using sophisticated, deep and exhaustive computational (either analytical and numerical) tools \cite{Cristini,Zheng}, we have exploited here simple, albeit effective, assumptions to find a link between Murray and Kleiber laws. We have indeed assumed that, at any node, the number of capillaries proliferates as $\dfrac{N_{k+1}}{N_{k}}=2$
while the relevant radius and length scale as $\dfrac{r_{k+1}}{r_{k}}=\dfrac{l_{k+1}}{l_{k}}=2^{-1}$.\\

The previous recurrences are not mandatory\footnote{We believe that some saturation mechanism should be introduced in the capillary branching process and the relevant proliferation recurrence should be replaced by $N_{k+1}=\dfrac{\alpha N_{k}}{1+\beta N_{k}}$ whose solution exhibits the logistic form $N_{k}=\dfrac{N_{0}\alpha^{k}}{1+\dfrac{N_{0}\beta}{\alpha -1}(\alpha^{k}-1)}$.}
neither is the growth factor $2$. It can be replaced by any other integer $n$. In the case of healthy tissues experimental evidence suggests, indeed, $n=3$ [see \cite{Savage} and referenes therein]\\

The problem of capillary vascularization in cancer has been formulated in even more general terms in ref. \cite{Savage}, where, albeit maintaining the assumption that, on the average the vascular network reproduces a self-similar fractal, the length and the radius exhibit the branching ''laws''
\begin{equation}\begin{split}\label{radLaws}
& \dfrac{r_{k+1}}{r_{k}}=n^{-\rho}\\
& \dfrac{l_{k+1}}{l_{k}}=n^{-\lambda}
\end{split}\end{equation}
Accordingly we find
\begin{equation}\begin{split}
& r_{k}=r_{0}n^{-\rho k}\\
& l_{k}=l_{0}n^{-\lambda k}
\end{split}\end{equation}
Eq. \ref{potenza}, detailing the network feeding power, should therefore be modified as
\begin{equation}
P_{T}=\alpha_{b}\pi \sum_{k=0}^{K}N_{k}r_{k}^{2}l_{k}
\end{equation}
No specific assumption has been made on the branching evolution of $N_{k}$ , however if we consider the volume as an invariant, namely
\begin{equation}
N_{k}r_{k}^{2}l_{k}=L_{0}r_{0}^{2}
\end{equation}
we obtain, on account of eqs. \ref{radLaws}, that
\begin{equation}
\dfrac{N_{k+1}}{N_{k}}=n^{2(\rho +\lambda)}
\end{equation}
Furthermore, by assuming that the Kleiber law holds, we obtain that the total number of capillaries satisfies the condition $N\propto V^{\frac{3}{4}}$.\\

This result is in substantial agreement with what has been obtained by Savage et al. \cite{Savage}, at least for the case of $\rho +\lambda >\frac{1}{2}$. Further comments can be found in that paper where an accurate description and analysis of capillary branching mechanism in healthy and cancer tissues has been developed.\\

We have identified macroscopic quantities, i.e. mass of the tumor and its radius, which can be exploited in a comparison with experimental or clinical data.
Most of the considerations contained in our paper, rely upon an evolution model based on the Verhulst equation or on its modified forms. Such a procedure is not new and, even though the literature in the field is considerable and authoritative and the assumptions underlying the models are convincing and well-motivated \cite{Agutter},  we believe that all of them exhibit significant limitations, if viewed as tools to analyze the experimental data. These models are all leading to the same mechanisms (growth and saturation) and differ only for some specific parameter (like the $\frac{3}{4}$ exponent in the Kleiber law), but, at least with regards to the analysis of the data, there is no clear evidence guiding the selection of one model versus the other. They are characterized by an S-shaped evolution of the cancer mass and, from the empirical point of view,  they are all equivalent. The various curves can indeed be mapped onto each other, by suitable rescaling of the parameters, thus hiding other effects, as shown in the following example. \\

The dormant pre-angiogenic phase or cell quiescence, induced by necrotic hypoxic signals and/or by systemic treatment,  rely on a largely unknown phenomenology, which can be accounted for in the evolution model only on the basis of ad hoc assumptions.
%\footnote{The term dormant may be source of confusion, it must be therefore clarified that the cells in these colonies have the same proliferation rate as in ''activ'' tumors, the difference being that their death rate is larger due to the lack or inefficiency of the associated capillary web. On the other side cells, inside a tumor, may become quiescent (in the sense that they reduce their reproduction activity) due to a variety of factors. The concepts of dormancy and quiescence should therefore be kept distinct}.\\

\begin{figure}[h]
          \centering
          \includegraphics[width=.4\textwidth]{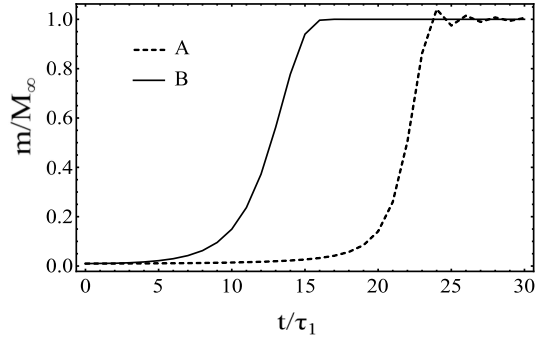}
          \caption{Cancer mass $\left(\frac{m}{M_{\infty}} \right)$ growth vs. time $\left(\frac{t}{\tau_{1}}\right)$ for the ordinary Verhulst and its modified form as reported in equation \ref{VerMc}; the graphs are obtained imposing the initial condition value ($\frac{m_{0}}{M_{\infty}}$) equal to $10^{-2}$ and with two different values for $M_{c}/M_{\infty}$ parameter: $\frac{M_{c}}{M_{\infty}}=10^{-1}$ for dashed line and $\frac{M_{c}}{M_{\infty}}=10^{-4}$ for continuous line.
                } \label{EvolRappDiff}
         \end{figure}

As already stressed, when the tumor undergoes the pre-angiogenetic and angiogenetic phases, it exhibits different growth behaviors. The evolution rate becomes exponential, as the number of cells increases. We can therefore introduce an empirical parameter (the critical mass $M_{c}$ associated e.g. with the largest mass compatible with an avascular growth), controlling this effect. Accordingly, we modify the Verhulst equation as follows \cite{Hassel}

\begin{equation}\begin{split}\label{VerMc}
& \dfrac{d}{dt}m=\dfrac{m}{\tau (m, M_{c})}\left(1-\dfrac{m}{M_{\infty}} \right) \\
& \tau (m, M_{c})=\dfrac{\tau}{\left(1-e^{-\frac{m}{M_{c}}} \right) }
\end{split}\end{equation}
The growth rate is modulated by the tumor mass itself and tends to approximate the constant value $\tau$ when $m>>M_{c}$, on the contrary when $m>>M_{c}$ the growth rate is significantly smaller. We expect therefore that the cancer mass evolution will take a larger amount of time to emerge from the initial latency stage to start its growth at an exponential rate. In Fig. \ref{EvolRappDiff}  we have reported the mass evolution predicted by eq. \ref{VerMc} for the ordinary Verhulst case while, for the case $F$ with
 $\dfrac{M_{c}}{M_{\infty}}=0.001$, the curve exhibit an almost identical behavior, even though shifted in time. Albeit such a time delay is expected, the use of one form or the other for the analysis of the data is not significant. The two curves can indeed be exactly overlapped after choosing different initial conditions.
 It is therefore evident that an increase of the latency time induced by a mechanism of this type could be inferred from the analysis of the experimental data only when initial conditions are well defined and this requires a very accurate control of the experimental
  conditions\footnote{An accurate control is extremely difficult because to be meaningful, clinical the data should be used as opposed to experimental in vivo or in vitro data.}.\\
%\begin{figure}[h]
%    \centering
%    \includegraphics[width=.4\textwidth]{evolutionConcluding.pdf}
%    \caption{\textbf{Cancer mass $\left(\dfrac{m}{M_{\infty}} \right)$ growth vs. $\left(\dfrac{t}{\tau_{1}} \right) $ for the ordinary Verhulst and its modified form \ref{VerMc} $\left(\dfrac{M_{c}}{M_{\infty}}=10^{-3}, \dfrac{m_{0}}{M_{\infty}}=10^{-4} \right)$
%}} \label{evolPrSec}
%   \end{figure}
The increase of the vascular feeding efficiency (associated with vascular sprouting and the consequent ability of the tumor to feed its cells more efficiently) may be even more effective and elude (within certain limits) the saturation mechanisms, by shifting the threshold of the on-set of the non-linearity. In other words it can be assumed that, when the mass increases, the system becomes capable of recruiting resources, thus acting as if the carrying capacity of the environment were increased by the growth of the tumor mass itself. \\

Such an effect is reminiscent of an analogous mechanism in social sciences regarding the human (population) evolution. That is, the carrying capacity of the ecological environment feeding humans seems to display a dynamical evolution, increasing with the population size itself.\\
%\footnote{his can be explained in different ways, including the technological innovations increasing the ability of finding further techniques for food resources. The present world-wide population could e. g. not survive by applying the agricultural technologies of the middle age.}.
The model, first proposed on the eve of the last century by Cohen \cite{Cohen}, assumes that the population growth is ruled by the following modified Verhulst equation
\begin{equation}\begin{split}\label{Cohen}
& \dfrac{d}{dt}m=\dfrac{m}{\tau_{1}}\left(1-\dfrac{m}{M_{\infty}(m, \lambda)} \right) \\
& \dfrac{d}{dt}M_{\infty}(m, \lambda)=\dfrac{\lambda}{m}\dfrac{d}{dt}m
\end{split}\end{equation}
In principle, this description could be used to model cancer cell evolution, by including new feeding strategies associated with vascularization and optimization of the resource recruiting mechanisms, through a dynamical tuning of the vascularweb architecture.\\

The use of equation (43) does not imply a departure from the logistic behavior, but only a shift in the saturation threshold, leaving unaffected the growth rate.\\

In this case too there is no way of selecting the effect from the analysis of the data of cancer mass evolution.
Both curves can indeed be interpolated with a Verhulst function differing only in the value of the saturated mass.\\

The same model can be adopted (just by reversing the sign of $\lambda$) to study effects induced by starvation of the tumor mass \cite{Freyer} therapy-induced or else.\\

\begin{figure}[h]
    \centering
    \includegraphics[width=.4\textwidth]{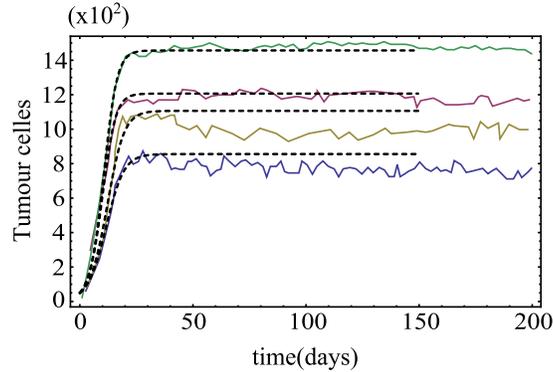}
    \caption{Comparison between the functions of eq. \ref{Cohen} (dashed line) and the experimental results (continuous line) for cancer cell evolution, with and without treatment (see text).
} \label{CnfrCurve}
   \end{figure}

Next, in Fig. \ref{CnfrCurve}
%e \ref{EvolRappDiff}
 we have therefore reported the result relevant to the evolution of cancer cells subject to different type of treatment \cite{Byrne}. The red line refers to the untreated evolution, while the others report the response to different treatments: immunotherapy (blue), chemotherapy (green), and a combination of both (black). Without focussing on the specific details we see that by assuming that the equation modelling the mass (number of cells) evolution is provided by eq. \ref{Cohen}, we can reproduce the same behavior after an appropriate choice of the parameter $\lambda$.\\

It is also worth noting that in the case of the green and black curves, a decrease of the malignant cell population follows the plateau phase; this effect too can be easily included in the model, as we will show elsewhere.

%   \begin{figure}[h]
%       \centering
%       \includegraphics[width=.4\textwidth]{EvolRappDiff.pdf}
%       \caption{\textbf{
%   }} \label{EvolRappDiff}
%      \end{figure}
The Verhulst evolution equation can be modified ad libitum, but it is our opinion, that no substantial improvement in our understanding of tumor behavior will be obtained through this modification strategy, if not implemented along the lines also discussed in \cite{Zheng}.\\

Arguably, the fractal nature of the capillary network in cancer is a main element of the cancer's survival strategy. \\

Furthermore, the distinction between primary and secondary tumors is perhaps inadequate. The fact that the cancer system sheds metastatic colonies at very early stages suggests that it develops relatively long term survival schemes, which are not only the consequence of subsistence needs due to environmental energetic crises.\\

Rather, the colonies may indeed stay dormant until the “primary” component is treated with radiation or chemo-therapies or  is surgically removed \cite{Hanin}. Sometime the impression one gets from this behavior is that  of a coherent system well integrated with the rest of the body. \\

We have stressed that the tumor undergoes the pre-angiogenetic and angiogenetic phases in which it exhibits different growth behaviors and becomes more efficient in finding energetic resources to increase its mass.
The success of the vascularization strategy stems from the “redundancy” of the process itself, namely from its ability of finding  different capillary web structures, which arguably is also the key tool to shield it from the impact of angio-genetic drugs  \cite{Hanin,Davies}. The ability of the cancer mass to establish advantageous links with the rest of the environment is somehow similar to what happens in the evolution of different types of populations or aggregates \cite{Batty}, based on allometry and agglomeration, size distribution and gravitational interaction.\\

Allometry is indeed a manifestation of the fractal nature of the geometry underlying the process, which governs the size. The gravitational interaction \cite{Mathur} is a concept borrowed from the physical language to model interaction between aggregates depending on their size and distance.\\

Geometry is an important, albeit scarcely considered, element in cancer evolution. The fractal nature of the process regulates the associated diffusion mechanisms, which cannot be modelled using ordinary diffusive equations of the heat type (albeit non-linear). We have emphasized the centrality of a Levy type process in cancer spreading and the fact that the relevant dynamics are regulated by fractional Fisher equations, in which the non-integer coefficient specifying the order of the involved derivatives is linked to the fractal exponent of the various scaling relations (such as Kleiber) involved in the macroscopic process that rules the evolution equation of the WBE type.\\

The emergency of necrotic cores in tumors is still a fact of noticeable importance \cite {Zheng} for its evolution strategy, but also the indication of a misconception of the associated geometrical structure. The topological idea we have in mind is that of a sphere and the core of the tumor is at center of this sphere.
 We have already stressed that the pathway of tumor expansion and capillary elongation occurs in opposite direction, but probably the geometrical concept of a center should be replaced with the “center” of energetic resources which lies outside the core of the tumor itself.\\

 This is not surprising, since increasing evidence suggests that many tumors are governed by the signals they receive from the microenvironment including the immune system cells. As such, cancer evolution and the aggressiveness of its metastasis is the result of  the interaction with signals inside and outside the tumor \cite{Davies}. And as such, hosting organ physiology and tumor should not be considered as separated entities.\\

%Body and tumor should therefore not considered as separated entities.\\
The last point of the analogy is what we have called gravitational interaction. A major interest is to better understand the interaction between the primary and secondary tumors which appears rather active and, as already stressed, dormant colonies become active when the main tumor is treated \cite{Hanin}.\\

We have here touched upon several central questions in cancer growth and disseminations and, supported by medical findings reported in the literature, have presented a set of innovative theoretical arguments. These build the basis for our future investigations in this important field.\\

{}

%\section{Appendix A}
%%\numberwithin{equation}{chapter}
%%\markboth{\textsc{\chaptername~\thechapter. Appendix A}}{}
%
%\section{Appendix B}
%%\numberwithin{equation}{chapter}
%%\markboth{\textsc{\chaptername~\thechapter. Appendix B}}{}

\end{document}